\documentstyle[12pt]{article}
\begin{document}
\title{SPHERICALLY SYMMETRIC DISSIPATIVE ANISOTROPIC FLUIDS: A GENERAL STUDY}
\author{L. Herrera$^1$\thanks{Postal
address: Apartado 80793, Caracas 1080A, Venezuela.} \thanks{e-mail:
laherrera@telcel.net.ve}, A. Di Prisco$^1$, J. Martin$^2$\thanks{email:chmm@usal.es}
, J. Ospino$^2$, 
N. O. Santos$^{3,4,5}$\thanks{e-mail: nos@cbpf.br; santos@ccr.jussieu.fr}\\ and  O. Troconis$^1$\\
{\small $^1$Escuela de F\'{\i}sica, Facultad de Ciencias,}\\
{\small Universidad Central de Venezuela, Caracas, Venezuela.} \\
{\small $^2$Area de F\'{\i}sica Te\'orica, Facultad de Ciencias,}\\ 
{\small Universidad de Salamanca, 37008 Salamanca, Espa\~na.}\\
{\small $^3$LERMA--UMR 8112--CNRS, ERGA Universit\'e PARIS VI,}\\
{\small  4 place Jussieu, 75005 Paris Cedex 05, France.}\\
{\small $^4$Laborat\'orio Nacional de Computa\c{c}\~ao Cient\'{\i}fica,}\\
{\small 25651-070 Petr\'opolis--RJ, Brazil.}\\
{\small $^5$ Centro Brasileiro de Pesquisas F\'{\i}sicas, 22290-180
Rio de Janeiro-RJ, Brazil.}
}
\maketitle
\newpage
\begin{abstract}
The full set of equations governing the evolution of self--gravitating spherically symmetric dissipative fluids with anisotropic stresses is deployed and used to carry out a general
study on the behaviour of such systems, in the context of general relativity.  Emphasis is given to the link  between the Weyl tensor, the shear tensor, the anisotropy of the
pressure and the density inhomogeneity. In particular we provide the general,  necessary and sufficient, condition for the vanishing of the spatial gradients of energy density, which in
turn suggests a possible definition of a gravitational arrow of time. Some solutions are also exhibited to illustrate the discussion.
\end{abstract}
\pagebreak
\section{Introduction}
This work is devoted to the study of dissipative, locally anisotropic, spherically symmetric self--gravitating fluids, with particular emphasis in a set of physical and geometrical
variables which  appear to play a fundamental role in the evolution of such systems. These variables are the Weyl tensor, the shear tensor, the local anisotropy of the presssure and the
density inhomogeneity. 

The Weyl tensor \cite{Pe} or some functions of it \cite{Wa}, have
been proposed to provide a gravitational arrow of time. The rationale
behind this idea being that tidal forces tend to make the gravitating
fluid more inhomogeneous as the evolution proceeds, thereby
indicating the sense of time. However, some works have thrown
doubts on this proposal \cite{arrow}. Further evidence about the relevance of the Weyl tensor in the evolution of self--gravitating systems may be found in \cite{HW}.

The role of density inhomogeneities in the collapse of dust \cite{MeTa}
and in particular in the formation of naked singularities \cite{VarI},
has been extensively discussed in the literature.

On the other hand, the assumption of local anisotropy of pressure, which seems to be very sensible to describe matter distribution under a variety of circumstances, has been proved
to be very useful in the study of relativistic compact objects
(see \cite{Lem} and references therein). 

A hint pointing to the relevance of the above mentioned three factors in
the fate of spherical collapse is also provided by the expression of
the active gravitational mass in terms of those factors \cite{HeSa95}.

Finally, the relevance of the shear tensor in the evolution of self--gravitaing systems has been brought out by many authors (see  \cite{shear} and references therein).

 Now, in the study of self--gravitating compact objects it is usually assumed that deviations from spherical
symmetry are likely to be incidental rather than  basic features of the process involved (see however the discussion in \cite{nonsph}). Thus, since  the seminal paper by
Oppenheimer and Snyder \cite{Opp}, most of the work dedicated to the problem of general relativistic gravitational collapse, deal with spherically symmetric fluid distribution.
Accordingly we shall consider spherically symmetric fluid distributions.

Also, the fluid distribution under consideration will be assumed to
be dissipative. Indeed, dissipation due to the emission of massless
particles (photons and/or neutrinos) is a characteristic process in the
evolution of massive stars. In fact, it seems that the only plausible
mechanism to carry away the bulk of the binding energy of the collapsing
star, leading to a neutron star or black hole is neutrino emission
\cite{1}. Consequently, in this paper, the matter distribution forming
the self--gravitating object will be described as a dissipative
fluid.

In the diffusion approximation, it is assumed that the energy flux of
radiation (as that of
thermal conduction) is proportional to the gradient of temperature. This
assumption is in general very sensible, since the mean free path of
particles responsible for the propagation of energy in stellar
interiors is in general very small as compared with the typical
length of the object.
Thus, for a main sequence star as the sun, the mean free path of
photons at the centre, is of the order of $2\, cm$. Also, the
mean free path of trapped neutrinos in compact cores of densities
about $10^{12} \, g.cm.^{-3}$ becomes smaller than the size of the stellar
core \cite{3,4}.

Furthermore, the observational data collected from supernova 1987A
indicates that the regime of radiation transport prevailing during the
emission process, is closer to the diffusion approximation than to the
streaming out limit \cite{5}.

However in many other circumstances, the mean free path of particles
transporting energy may be large enough as to justify the  free streaming
approximation. Therefore we
shall include simultaneously both limiting  cases of radiative transport
(diffusion and streaming out), allowing for describing a wide range
situations.

It is also worth
mentioning that although the most common method of solving Einstein's
equations is to use commoving coordinates
(e.g.
\cite{May}), we shall use noncomoving coordinates, which implies that
the velocity of any fluid element (defined with respect to a conveniently
chosen set of observers) has to be
considered as a relevant physical variable (\cite{Knutsen}). 

The paper is organized as follows: In the next section we introduce the notation and write all relevant equations. Section 3 is devoted to the analysis of different special 
cases. Finally the results are discussed in the  last section.

\section{The basic equations}
 
In this section we shall deploy the relevant equations for describing a dissipative self--gravitating locally anisotropic fluid. In spite of the fact that not all these equations are
independent (for example the field equations and the conservation equations (Bianchi identities)) we shall present them all, since depending on the problem under consideration, it may
be more advantageous using one set instead of the other.

\subsection{Einstein equations}

We consider spherically symmetric distributions of collapsing
fluid, which for sake of completeness we assume to be locally anisotropic,
undergoing dissipation in the form of heat flow and/or free streaming
radiation, bounded by a
spherical surface $\Sigma$.

\noindent
The line element is given in Schwarzschild--like coordinates by

\begin{equation}
ds^2=e^{\nu} dt^2 - e^{\lambda} dr^2 -
r^2 \left( d\theta^2 + sin^2\theta d\phi^2 \right),
\label{metric}
\end{equation}

\noindent
where $\nu(t,r)$ and $\lambda(t,r)$ are functions of their arguments. We
number the coordinates: $x^0=t; \, x^1=r; \, x^2=\theta; \, x^3=\phi$.

\noindent
The metric (\ref{metric}) has to satisfy Einstein field equations

\begin{equation}
G^\nu_\mu=8\pi T^\nu_\mu,
\label{Efeq}
\end{equation}

\noindent
which in our case read \cite{Bo}:

\begin{equation}
-8 \pi T^0_0 =-{1 \over r^2} +e^{-\lambda} ({1 \over r^2}
-{\lambda ^{\prime} \over r})
\label{1}
\end{equation}
\begin{equation}
-8 \pi T^1_1=-{1 \over r^2} +e^{-\lambda} ({1 \over r^2} +{{\nu
^{\prime}} \over r})
\label{2}
\end{equation}

$$-8 \pi T^2_2=-8 \pi T^3_3=-\frac{e^{-\nu}}{4} (2 \ddot
\lambda+\dot \lambda  (\dot \lambda -\dot \nu ))$$
\begin{equation}
+\frac{e^{-\lambda}}{4}(2 \nu ^{\prime \prime} + {\nu
^{\prime}}^2-\lambda ^{\prime} \nu ^{\prime} + 2 \frac{\nu
^{\prime}-\lambda ^{\prime}}{r})
\label{3'}
\end{equation}
\begin{equation}
-8 \pi T_{10} =-\frac{\dot \lambda}{r},
\label{4}
\end{equation}
\noindent
where dots and primes stand for partial differentiation with respect
to $t$ and $r$,
respectively.

\noindent
In order to give physical significance to the $T^{\mu}_{\nu}$ components
we apply the Bondi approach \cite{Bo}.

\noindent
Thus, following Bondi, let us introduce purely locally Minkowski
coordinates ($\tau, x, y, z$)

$$d\tau=e^{\nu/2}dt\,;\qquad\,dx=e^{\lambda/2}dr\,;\qquad\,
dy=rd\theta\,;\qquad\, dz=rsin\theta d\phi.$$

\noindent
Then, denoting the Minkowski components of the energy tensor by a bar,
we have

$$\bar T^0_0=T^0_0\,;\qquad\,
\bar T^1_1=T^1_1\,;\qquad\,\bar T^2_2=T^2_2\,;\qquad\,
\bar T^3_3=T^3_3\,;\qquad\,\bar T_{01}=e^{-(\nu+\lambda)/2}T_{01}.$$

\noindent
Next, we suppose that when viewed by an observer moving relative to these
coordinates with proper velocity $\omega$ in the radial direction, the physical
content  of space consists of an anisotropic fluid of energy density $\rho$,
radial pressure $P_r$, tangential pressure $P_\bot$,  radial heat flux
$q$ and unpolarized radiation of energy density $\epsilon$
traveling in the radial direction. Thus, when viewed by this moving
observer the covariant tensor in
Minkowski coordinates is

\[ \left(\begin{array}{cccc}
\rho + \epsilon    &  - q -\epsilon  &   0     &   0    \\
- q - \epsilon &  P_r + \epsilon    &   0     &   0    \\
0       &   0       & P_\bot  &   0    \\
0       &   0       &   0     &   P_\bot
\end{array} \right). \]

\noindent
Then a Lorentz transformation readily shows that
\begin{equation}
T^0_0= \bar T^0_0 =\frac{\rho+P_r \omega ^2}{1-\omega
^2}+\frac{2\omega
q}{1-\omega^2}+\frac{\epsilon(1+\omega)}{1-\omega}
\label{8'}
\end{equation}
\begin{equation}
T^1_1=\bar T^1_1 =-\frac{P_r+\rho \omega ^2}{1-\omega
^2}-\frac{2\omega
q}{1-\omega^2}-\frac{\epsilon(1+\omega)}{1-\omega}
\label{9}
\end{equation}
\begin{equation}
T^2_2=T^3_3=\bar T^2_2 =\bar T^3_3 =-P_{\bot}
\label{10}
\end{equation}
\begin{equation}
T_{01}=e^{\frac{\nu+\lambda}{2}}\bar
T_{01}=-\frac{(\rho+P_r)\omega e^{\frac{\nu+\lambda}{2}}}{1-\omega
^2}-\frac{q e^{\frac{\lambda+\nu}{2}}}{1-\omega ^2}(1+\omega
^2)-\frac{e^{\frac{\lambda +\nu}{2}}
\epsilon(1+\omega)}{1-\omega}
\label{11a}
\end{equation}

\noindent
Note that the coordinate velocity in the ($t,r,\theta,\phi$) system, $dr/dt$,
is related to $\omega$ by

\begin{equation}
\omega=\frac{dr}{dt}\,e^{(\lambda-\nu)/2}.
\label{omega}
\end{equation}

\noindent
Feeding back (\ref{8'}--\ref{11a}) into (\ref{1}--\ref{4}), we get
the field equations in  the form

\begin{equation}
\frac{\rho + P_r \omega^2 }{1 - \omega^2} +\frac{2\omega
q}{1-\omega^2}+\frac{\epsilon(1+\omega)}{1-\omega}=-\frac{1}{8 \pi}\Biggl\{-\frac{1}{r^2}+e^{-\lambda}
\left(\frac{1}{r^2}-\frac{\lambda'}{r} \right)\Biggr\},
\label{fieq00}
\end{equation}

\begin{equation}
\frac{ P_r + \rho \omega^2}{1 - \omega^2} +
\frac{2\omega
q}{1-\omega^2}+\frac{\epsilon(1+\omega)}{1-\omega}=-\frac{1}{8 \pi}\Biggl\{\frac{1}{r^2} - e^{-\lambda}
\left(\frac{1}{r^2}+\frac{\nu'}{r}\right)\Biggr\},
\label{fieq11}
\end{equation}

\begin{eqnarray}
P_\bot = -\frac{1}{8 \pi}\Biggl\{\frac{e^{-\nu}}{4}\left(2\ddot\lambda+
\dot\lambda(\dot\lambda-\dot\nu)\right) \nonumber \\
 - \frac{e^{-\lambda}}{4}
\left(2\nu''+\nu'^2 -
\lambda'\nu' + 2\frac{\nu' - \lambda'}{r}\right)\Biggr\},
\label{fieq2233}
\end{eqnarray}

\begin{equation}
\frac{(\rho + P_r) \omega e^{\frac{\lambda +\nu}{2}}}{1 - \omega^2} +\frac{q e^{\frac{\lambda+\nu}{2}}}{1-\omega ^2}(1+\omega
^2)+\frac{e^{\frac{\lambda +\nu}{2}}
\epsilon(1+\omega)}{1-\omega}=-\frac{\dot\lambda}{8 \pi r}.
\label{fieq01}
\end{equation}

The four--velocity  vector is defined as
\begin{equation}
u^{\alpha}=(\frac{e^{-\frac{\nu}{2}}}{(1-\omega^2)^{\frac{1}{2}}},\frac{\omega
e^{ -\frac{\lambda}{2}}}{(1-\omega^2)^{\frac{1}{2}}},0,0).
\end{equation}
 From which we can calculate the four acceleration
$a^\alpha=u^\alpha_{;\beta}u^\beta$ to obtain
\begin{equation}
\omega a_1=-a_0 e^{\frac{\lambda-\nu}{2}}=-\frac{\omega}{1-\omega
^2} [(\frac{\omega \omega ^{\prime}}{1-\omega^2}+\frac{\nu
^{\prime}}{2})+e^{\frac{\lambda-\nu}{2}}(\frac{\omega \dot
\lambda}{2}+\frac{\dot \omega}{1-\omega^2})].
\label{15a}
\end{equation}

\noindent
For the exterior of the fluid distribution, the spacetime is that of Vaidya,
given by

\begin{equation}
ds^2= \left(1-\frac{2M(u)}{R}\right) du^2 + 2dudR -
R^2 \left(d\theta^2 + sin^2\theta d\phi^2 \right),
\label{Vaidya}
\end{equation}

\noindent
where $u$ is a coordinate related to the retarded time, such that
$u=constant$ is (asymptotically) a
null cone open to the future and $R$ is a null coordinate ($g_{
RR}=0$). 
\noindent

The two coordinate systems ($t,r,\theta,\phi$) and ($u,
R,\theta,\phi$) are
related at the boundary surface and outside it by

\begin{equation}
u=t-r-2M\,ln \left(\frac{r}{2M}-1\right),
\label{u}
\end{equation}

\begin{equation}
{R}=r.
\label{radial}
\end{equation}

\noindent
In order to match smoothly the two metrics above on the boundary surface
$r=r_\Sigma(t)$, we  require the continuity of the first and the second fundamental
forms across that surface, yielding (see \cite{HJR} for details)

\begin{equation}
e^{\nu_\Sigma}=1-\frac{2M}{R_\Sigma},
\label{enusigma}
\end{equation}
\begin{equation}
e^{-\lambda_\Sigma}=1-\frac{2M}{R_\Sigma},
\label{elambdasigma}
\end{equation}
\begin{equation}
\left[P_r\right]_\Sigma=\left[q \right]_\Sigma .
\label{PQ}
\end{equation}
Where, from now on, subscript $\Sigma$ indicates that the quantity is
evaluated on the boundary surface $\Sigma$, and (\ref{PQ}) expresses the discontinuity of the radial pressure in the presence
of heat flow, which is a well known result \cite{Sa}.

Eqs. (\ref{enusigma}), (\ref{elambdasigma}), and (\ref{PQ}) are the necessary and
sufficient conditions for a smooth matching of the two metrics (\ref{metric})
and (\ref{Vaidya}) on $\Sigma$.

\subsection{Conservation laws ($T^{\mu}_{\nu;\mu}=0$)}

The energy--momentum tensor  (\ref{8'})-(\ref{11a}) may be
written as :
\begin{equation}
T^{\mu}_{\nu}=\tilde \rho u^{\mu}u_{\nu}- \hat P
h^{\mu}_{\nu}+\Pi ^{\mu}_{\nu} +\tilde q(s^\mu u_\nu+s_\nu u^\mu),
\label{24'}
\end{equation}
with $$h^{\mu}_{\nu}=\delta^{\mu}_{\nu}-u^{\mu}u_{\nu},$$ $$\Pi
^{\mu}_{\nu}=\Pi(s^{\mu}s_{\nu}+\frac{1}{3}h^{\mu}_{\nu}),$$
$$\hat P=\frac{\tilde P_{r}+2P_{\bot}}{3},$$

$$\tilde \rho= \rho+\epsilon,$$
$$\tilde P_{r}=P_r+\epsilon,$$
$$\tilde q= q+\epsilon,$$
$$\Pi=\tilde P_{r}-P_{\bot}$$
and  $s^\mu$ is defined as
\begin{equation}
s^{\mu}=(\frac{\omega
e^{-\frac{\nu}{2}}}{(1-\omega^2)^{\frac{1}{2}}},
\frac{e^{-\frac{\lambda}{2}}}{(1-\omega^2)^{\frac{1}{2}}},0,0)
\end{equation}

with the  properties
$s^{\mu}u_{\mu}=0$,
$s^{\mu}s_{\mu}=-1$, and $\tilde q^{\mu}=\tilde qs^{\mu}$.

 We may write for the shear tensor 
\begin{equation}
 \sigma _{\alpha \beta}=\frac{1}{2}\sigma (s_\alpha
s_\beta+\frac{1}{3}h_{\alpha_\beta})
\label{52}
\end{equation}

with

\begin{equation}
\sigma=-\frac{1}{(1-\omega^2)^{\frac{1}{2}}}[e^{-\frac{\nu}{2}}(\dot
\lambda+\frac{2\omega \dot \omega}{1-\omega
^2})+e^{-\frac{\lambda}{2}}(\omega \nu ^{\prime}+\frac{2\omega
^{\prime}}{1-\omega ^2}-\frac{2\omega}{r})].
\label{sigma}
\end{equation}

Then from $T^{\mu}_{\nu;\mu}=0$, using (\ref{24'}), we find:
\begin{equation}
\tilde \rho_{;\alpha} u^{\alpha} +(\tilde \rho+\hat P)\theta+\tilde
q^{\alpha}_{;\alpha}=\Pi_{\alpha\beta} \sigma
^{\alpha\beta}+\tilde q a^\nu s_\nu
\label{26'}
\end{equation}

and

\begin{equation}
(\tilde \rho+\hat P) a_\alpha+h^{\beta}_{\alpha}(\tilde 
q_{;\nu}u^{\nu} s_{\beta}+\tilde q  s_{\beta;\nu}u^{\nu}-\hat 
{P}_{,\beta}+\Pi ^{\mu}_{\beta
;\mu})+\sigma_{\alpha\beta}\tilde
qs^{\beta}+\frac{4}{3}\theta \tilde qs_{\alpha}=0.
\label{27'}
\end{equation}
Or, contracting  (\ref {27'}) with  $s^\alpha $:
$$\tilde P_{r;\mu}s^\mu+(\tilde P_r-P_\bot)s^\mu_{;\mu}-(\tilde \rho
+P_\bot)a_\mu s^\mu+\frac{4}{3}\theta \tilde q+\tilde
q_{;\nu}u^{\nu}-\tilde qs^\mu s^\nu \sigma_{\mu \nu}=0.$$

\subsection{Ricci identities}
  Ricci identitites for the vector  $u_\alpha$
read:
\begin{equation}
u_{\alpha ;\beta ; \nu}- u_{\alpha;\nu;\beta}=R^{\mu}_{\alpha
\beta \nu}u_\mu ,
\end{equation}
or using
\begin{equation}
u_{\alpha;\beta}=a_\alpha u_\beta +\sigma_{\alpha
\beta}+\frac{1}{3}\theta h_{\alpha \beta},
\label{29}
\end{equation}
we have 
\begin{equation}
\frac{1}{2}R^{\rho}_{\alpha \beta
\mu}u_{\rho}=a_{\alpha;[\mu}u_{\beta]}+a_{\alpha}u_{[\beta;\mu]}+\sigma_{\alpha
[\beta;\mu]}+\frac{1}{3}\theta_{,[\mu}h_{\beta]\alpha}+\frac{1}{3}\theta
h_{\alpha[\beta;\mu]}.
\label{30}
\end{equation}
\subsubsection{Raychaudhuri equation}
Contracting (\ref{30}) with $u^\beta$ and then the indices $\alpha $ and  $\mu$,
 we find the Raychaudhuri equation for the evolution of the expansion:
\begin{equation}
\theta_{;\alpha}u^{\alpha} +\frac{\theta 
^2}{3}+\sigma_{\alpha\beta}\sigma^{\alpha\beta}
-a^{\alpha}_{;\alpha}=-u_\rho u^{\beta}R^\rho_\beta=-4\pi(\tilde
\rho+3\hat P),
\label{31}
\end{equation}
where
$$
\sigma_{\alpha\beta}\sigma^{\alpha\beta}=\frac{1}{6} \sigma ^2
.$$

\subsubsection{Constraint equation}
If in  (\ref{30}) we contract first  $\alpha $ and
$\mu$ and then contract with $h^{\alpha \beta}$,
we obtain the constraint equation expressing a direct relation
between expansion
$\theta$, shear
$\sigma ^{\alpha
\beta}$ and the heat flux $q$:
\begin{equation}
R^\rho_\beta u_\rho h^{\alpha \beta}
=h^{\alpha}_{\beta}(\sigma^{\beta
\mu}_{;\mu}-\frac{2}{3}\theta ^{;\beta})+\sigma^{\alpha
\beta}a_{\beta}=8\pi \tilde qs^{\alpha}.
\label{32}
\end{equation}

\subsubsection{Propagation equation of the shear}
 Contracting (\ref{30}) with
$u^{\beta} h^{\alpha}_\gamma h^\mu_\nu$ we have
\begin{equation}
u_\rho u^\beta R^\rho_{\alpha \beta\mu}h^\alpha_\gamma h^\mu_\nu =
h^\alpha_\gamma h^\mu_\nu(a_{\alpha;\mu}-
\sigma_{\alpha\mu;\beta}u^{\beta})-a_\gamma
a_\nu-u^\beta_{;\mu}h^\mu_\nu(\sigma_{\gamma
\beta}+\frac{\theta}{3}h_{\gamma 
\beta})-\frac{\theta_{;\alpha}u^{\alpha}}{3}h_{\gamma \nu}.
\label{33}
\end{equation}
On the other hand we know that the
Riemann tensor may be expressed through the Weyl tensor
$C^{\rho}_{\alpha
\beta
\mu}$, the  Ricci tensor $R_{\alpha\beta}$ and the scalar curvature $R$,
as:
$$
R^{\rho}_{\alpha \beta \mu}=C^\rho_{\alpha \beta \mu}+ \frac{1}{2}
R^\rho_{\beta}g_{\alpha \mu}-\frac{1}{2}R_{\alpha \beta}\delta
^\rho_{\mu}+\frac{1}{2}R_{\alpha \mu}\delta^\rho_\beta$$
\begin{equation}
-\frac{1}{2}R^\rho_\mu g_{\alpha
\beta}-\frac{1}{6}R(\delta^\rho_\beta g_{\alpha \mu}-g_{\alpha
\beta}\delta^\rho_\mu).
\label{34}
\end{equation}
Contracting  (\ref{34}) with$ u_{\rho} u^\beta
h^\alpha_\gamma h^\mu_\nu$ and using  Einstein equation
(\ref{Efeq}), we find:
\begin{equation}
R^\rho_{\alpha \beta \mu}u_{\rho} u^\beta
h^\alpha_\gamma h^\mu_\nu=E_{\gamma \nu}+4\pi \Pi
_{\gamma \nu}+\frac{4\pi}{3}h_{\gamma \nu}(\tilde \rho+3\hat P),
\label{35}
\end{equation}
where $E_{\gamma \nu}$ denotes the ``electric'' part of the Weyl tensor defined  by the equation (\ref{40}) below.

From (\ref{33}) and  (\ref{35}) taking into account (\ref{31}) it follows:
\begin{equation}
E_{\gamma \nu}+4\pi \Pi_{\gamma \nu}=h^\alpha _\gamma h^\mu 
_\nu(a_{\alpha;\mu}-
\sigma_{\alpha \mu;\beta}u^{\beta})-a_\gamma a_\nu-\sigma^\beta_\nu
\sigma_{\gamma \beta}-\frac{2}{3}\theta \sigma_{\gamma 
\nu}-\frac{1}{3}(a^\alpha
_{; \alpha}-\frac{1}{6} \sigma^2 )h_{\gamma \nu}
\label{36}
\end{equation}

\subsection{Evolution equations for the  Weyl tensor} 
According to  Kundt and Tr\"{u}mper \cite{Kundt}, Bianchi identitites
\begin{equation}
R_{\mu\nu \kappa \delta ;\lambda}+R_{\mu \nu \lambda \kappa ;\delta}+R_{\mu
\nu \delta \lambda;\kappa}=0,
\end{equation}
may be written as:
\begin{equation}
C^{\,\,\,\, \,\, \,\,\,\, \lambda}_{\mu \nu \kappa\,\,\,
;\lambda}=R_{\kappa[\mu;\nu]}-\frac{1}{6}g_{\kappa[\mu}R_{,\nu]}.
\label{38}
\end{equation}
Then taking into account Einstein equations (\ref{Efeq}),
  (\ref{38}) reads:
\begin{equation}
C^{\,\,\,\, \,\, \,\,\,\, \lambda}_{\mu \nu \kappa\,\,\,
;\lambda}=8\pi T_{\kappa[\mu;\nu]}-\frac{8\pi}{3}g_{\kappa[\mu}T_{,\nu]}.
\label{39}
\end{equation}

In the spherically symmetric case the ``magnetic'' part of the Weyl tensor
vanishes($H_{\alpha \beta}=0$), then we have:
\begin{equation}
C_{\mu \nu \kappa \lambda}=(g_{\mu\nu \alpha \beta}g_{\kappa \lambda \gamma
\delta}-\epsilon_{\mu\nu \alpha \beta}\epsilon_{\kappa \lambda \gamma
\delta})u^\alpha u^\gamma E^{\beta \delta},
\label{40}
\end{equation}
with $g_{\mu\nu \alpha \beta}=g_{\mu \alpha}g_{\nu \beta}-g_{\mu
\beta}g_{\nu \alpha}$,  $\epsilon_{\mu\nu \alpha \beta}$
is the Levi-Civita symbol multiplied by $\sqrt{-g}$ and  $E^{\beta 
\gamma}$, the ``electric'' part of
Weyl tensor, may be written as:
\begin{equation}
E_{\alpha \beta}=E (s_\alpha s_\beta+\frac{1}{3}h_{\alpha \beta})
\label{52}
\end{equation}
with
\begin{eqnarray}
E=\frac{e^{-\nu}}{4} ( \ddot
\lambda+\frac{\dot \lambda (\dot \lambda -\dot \nu )}{2})\\ \nonumber
\\ \nonumber
-\frac{e^{-\lambda}}{4}( \nu ^{\prime \prime} + \frac{{\nu
^{\prime}}^2-\lambda ^{\prime} \nu ^{\prime}}{2} -  \frac{\nu
^{\prime}-\lambda ^{\prime}}{r}+\frac{2(1-e^{\lambda})}{r^2})
\end{eqnarray}

Contracting  (\ref{40}) with $u^\nu$ we obtain:
\begin{equation}
u^\nu C_{\mu \nu k \lambda}=E_{\mu k}u_{\lambda}-E_{\mu
\lambda}u_k,
\end{equation}
from where it follows that:
\begin{equation}
u^\nu C^{\,\,\,\,\,\,\,\,\, \lambda} _{\mu \nu \kappa\, \, \, ;
\lambda}+u^\nu_{;\lambda}C^{\,\,\,\,\,\,\,\,\, \lambda}_{\mu\nu
\kappa}=\theta E_{\mu \kappa}+u^{\alpha} E_{\mu \kappa;\alpha} 
-u_{\kappa;\lambda}E^\lambda_\mu -
u_\kappa E^\lambda _{\mu; \lambda}.
\label{42}
\end{equation}
Also, from (\ref{29}) and (\ref{40}), we obtain:

\begin{equation}
u^\nu_{;\lambda}C^{\,\,\,\,\,\,\,\,\, \lambda}_{\mu\nu
\kappa}=u_\mu u_\kappa\sigma_{\delta \beta} E^{\delta
\beta}-a_\beta u_\mu E^\beta_\kappa-h_{\mu \kappa}\sigma^{\alpha 
\beta} E_{\alpha
\beta}+\sigma_{\kappa \alpha}E^{\alpha}_{\mu}+ \sigma_{\mu
\alpha}E^{\alpha}_{\kappa}-\frac{\theta}{3}E_{\mu \kappa}.
\label{43}
\end{equation}
Replacing (\ref{43}) into (\ref{42}), it results:
\begin{eqnarray}
u^\nu C^{\,\,\,\,\,\,\,\,\, \lambda} _{\mu \nu \kappa\, \, \, ;
\lambda}=& &\frac{4\theta}{3} E_{\mu \kappa}+u^{\alpha} E_{\mu \kappa;\alpha}
-u_{\kappa;\lambda}E^\lambda_\mu-u_\kappa E^\lambda _{\mu;
\lambda}-u_\mu u_\kappa \sigma_{\delta \beta} E^{\delta
\beta}+ \nonumber \\
&+&a_\beta u_\mu E^\beta_\kappa  
+h_{\mu \kappa}\sigma^{\alpha \beta}\
E_{\alpha
\beta}-\sigma_{\kappa \alpha}E^{\alpha}_{\mu}- \sigma_{\mu
\alpha}E^{\alpha}_{\kappa}.
\label{44a}
\end{eqnarray}
Contracting (\ref{44a}) with $h^\mu _\alpha \, h^\kappa_\beta$ we have:
\begin{eqnarray}
h^\mu_\alpha \, h^\kappa_\beta \,u^\nu C^{\,\,\,\,\,\,\,\,\, \lambda} _{\mu \nu
\kappa\, \, \, ; \lambda}&=&\frac{4\theta}{3}
E_{\alpha \beta}-u_{\beta;\lambda}E^\lambda _\alpha+u^{\nu} E_{\mu \kappa;\nu}
h^\mu_\alpha\, h^\kappa_\beta + \nonumber \\
& +&h_{\alpha \beta}\sigma^{\kappa \nu}\
E_{\kappa
\nu} 
-\sigma_{\kappa \alpha}E^{\kappa}_{\beta}- \sigma_{\kappa
\beta}E^{\kappa}_{\alpha}.
\label{45}
\end{eqnarray}
On the other hand
\begin{equation}
\left. \begin{array}{l}
h^\mu_\alpha \, h^\kappa_\beta \, u^\nu \, T_{\kappa \mu ;
\nu}=-u^{\nu}{\hat P_{;\nu}}h_{\alpha \beta}+u^{\nu} \Pi _{\kappa 
\mu;\nu}h^\mu _\alpha h^\kappa_\beta+\tilde{q}_\alpha a_\beta
+\tilde{q}_\beta a_\alpha
\\
\\
h^\mu_\alpha \, h^\kappa_\beta u^\nu T_{\kappa \nu ; \mu}=(\tilde \rho +\hat
P)(\sigma_{\alpha \beta}+\frac{\theta}{3}h_{\alpha \beta})-\Pi_{\beta
\nu}(\sigma ^\nu_\alpha +
\frac{\theta}{3}h^\nu_\alpha)+{\tilde q}_{\kappa;\mu}\,h^\mu_\alpha 
h^\kappa_\beta
\\
\\
h^\mu_\alpha \, h^\kappa_\beta \, u^\nu \, g_{\kappa 
[\mu}T_{,\nu]}=\frac{1}{2}u^{\nu}(
{\tilde \rho_{;\nu}}-3{\hat P_{;\nu}})h_{\alpha \beta}
\end{array} \right \}
\label{46}
\end{equation}
Feeding back (\ref{45}) and (\ref{46}) into (\ref{39}) we find:
 \begin{eqnarray}
\theta
E_{\alpha \beta}+(u^{\nu} E_{\mu
\kappa;\nu}-4\pi u^{\nu}\Pi_{\mu \kappa;\nu}+ 4\pi \tilde 
q_{\kappa;\mu})h^\mu_\alpha h^\kappa_\beta
+\frac{4\pi}{3}u^{\nu}{\tilde \rho_{;\nu}}h_{\alpha \beta}+
E\sigma_{\alpha \beta}=
\\ \nonumber
-4\pi (\tilde \rho+\hat
P)(\sigma_{\alpha \beta}+\frac{\theta}{3} h_{\alpha \beta})+4\pi (\tilde
q_\alpha a_\beta+\tilde q_\beta a_\alpha) +4\pi \Pi _{\nu
\beta}(\sigma^\nu_\alpha+\frac{\theta}{3}h^\nu _\alpha)
\label{47}
\end{eqnarray}

Next, contracting (\ref{44a}) with $u^k$ we have:

\begin{equation}
u^\kappa u^\nu C^{\,\,\,\,\,\,\,\,\, \lambda} _{\mu \nu k\, \, \, ;
\lambda}=-E^{\lambda}_{\mu ; \lambda}-a^\lambda E_{\mu
\lambda}-\sigma ^\nu _{\lambda}E^\lambda_\nu u_\mu.
\label{48}
\end{equation}
The following expressions can also be easily calculated:
\begin{equation}
\left. \begin{array}{l} u^k  \, u^\nu T_{k \mu ; \nu}=u^{\nu}{\tilde
\rho_{;\nu}}u_\mu+(\tilde \rho+\hat P)a_\mu-a^k(\tilde
q_ku_\mu+\Pi_{\mu k})+u^{\nu}{\tilde q_{\mu;\nu}}
\\
\\
u ^k \, u^\nu  T_{k \nu ; \mu}=\tilde \rho_{;\mu}-2\tilde
q_k(a^ku_\mu+ \sigma ^k_\mu+\frac{\theta}{3}h^k_\mu)
\\
\\
g_{k[\mu}T_{;\nu]}u^\nu u^k=-\frac{1}{2}T_{,\nu}
h^\nu_\mu=-\frac{1}{2}(\tilde \rho -3\tilde P)_{,\nu}h^\nu_\mu
\end{array} \right \}
\label{49}
\end{equation}
Finally, feeding back (\ref{48}) and (\ref{49}) into (\ref{39}) and contracting with $h^\mu
_\alpha$  we have:
\begin{eqnarray}
  E^\lambda_{\mu;\lambda}h^\mu_\alpha+a^\lambda
E_{\alpha \lambda}=-8\pi \tilde
q_\kappa(\sigma^\kappa_\alpha+\frac{\theta}{3}h^\kappa_\alpha)+\frac{4\pi}{3}(2
\tilde \rho
+3\hat P)_{; \nu}h^\nu_\alpha \\ \nonumber
\\ \nonumber
\,\,\,\,\,\,\,\,\,\, \,\,\,\,\,\,\,\,\,\,\,\,\,\,\,\,\,\,\,\,
\,\,\,\,\,\,\,\,\,\, -4\pi(\tilde \rho+\hat P)a_\alpha+4\pi a^\kappa
\Pi_{\alpha \kappa}-4\pi u^{\nu} {\tilde q_{\mu;\nu}}h^\mu_\alpha
\label{50}
\end{eqnarray}

\subsection{Weyl tensor, mass function  and anisotropy}
For the line element (\ref{metric}) we have:
\begin{equation}
R^3_{232}=1-e^{-\lambda}=\frac{2m}{r}
\label{rieman}
\end{equation}
Where the mass function $m(r,t)$ is defined as
\begin{equation}
m = 4 \pi \int^{r}_{0}{r^2 T^0_0 dr}, \\
\label{m}
\end{equation}
Then from (\ref{34}), (\ref{52}) and  Einstein equations (\ref{Efeq}) it follows:
\begin{equation}
\frac{3m}{r^3}=4\pi \tilde \rho +4\pi (P_\bot -\tilde P_r)+E
\label{66}                   
\end{equation}
which in tensorial form reads
\begin{equation}
E_{\alpha \beta}-4\pi \Pi _{\alpha \beta} =(\frac{3m}{r^3}-4\pi
\tilde \rho)(s_\alpha s_\beta +\frac{1}{3}h_{\alpha \beta})
\label{66t}
\end{equation}

\subsection{Summary}
Equations  (\ref{26'}), (\ref{27'}), (\ref{31}), (\ref{32}), (\ref{36}), (\ref{47}), (\ref{50}) and (\ref{66t}) are
\begin{equation}
\tilde \rho_{;\alpha} u^{\alpha} +(\tilde \rho+\hat P)\theta+\tilde
q^{\alpha}_{;\alpha}=\Pi_{\alpha\beta} \sigma
^{\alpha\beta}+\tilde q a^\nu s_\nu
\label{a}
\end{equation}
\\

\begin{equation}
(\tilde \rho+\hat P) a_\alpha+h^{\beta}_{\alpha}(\tilde 
q_{;\nu}u^{\nu} s_{\beta}+\tilde q  s_{\beta;\nu}u^{\nu}-\hat 
{P}_{,\beta}+\Pi ^{\mu}_{\beta
;\mu})+\sigma_{\alpha\beta}\tilde
qs^{\beta}+\frac{4}{3}\theta \tilde qs_{\alpha}=0
\label{b}
\end{equation}
\\
\begin{equation}
\theta_{;\alpha}u^{\alpha} +\frac{\theta 
^2}{3}+\sigma_{\alpha\beta}\sigma^{\alpha\beta}
-a^{\alpha}_{;\alpha}=-u_\rho u^{\beta}R^\rho_\beta=-4\pi(\tilde
\rho+3\hat P)
\label{c}
\end{equation}
\\
\begin{equation}
R^\rho_\beta u_\rho h^{\alpha \beta}
=h^{\alpha}_{\beta}(\sigma^{\beta
\mu}_{;\mu}-\frac{2}{3}\theta ^{;\beta})+\sigma^{\alpha
\beta}a_{\beta}=8\pi \tilde qs^{\alpha}
\label{d}
\end{equation}
\\
\begin{equation}
E_{\gamma \nu}+4\pi \Pi_{\gamma \nu}=h^\alpha _\gamma h^\mu 
_\nu(a_{\alpha;\mu}-
\sigma_{\alpha \mu;\beta}u^{\beta})-a_\gamma a_\nu-\sigma^\beta_\nu
\sigma_{\gamma \beta}-\frac{2}{3}\theta \sigma_{\gamma 
\nu}-\frac{1}{3}(a^\alpha
_{; \alpha}-\frac{1}{6} \sigma^2 )h_{\gamma \nu}
\label{e}
\end{equation}
\\
\begin{eqnarray}
 \theta
E_{\alpha \beta}+(u^{\nu} E_{\mu
\kappa;\nu}-4\pi u^{\nu}\Pi_{\mu \kappa;\nu}+ 4\pi \tilde 
q_{\kappa;\mu})h^\mu_\alpha h^\kappa_\beta
+\frac{4\pi}{3}u^{\nu}{\tilde \rho_{;\nu}}h_{\alpha \beta}+
E\sigma_{\alpha \beta}=
\\ \nonumber
-4\pi (\tilde \rho+\hat
P)(\sigma_{\alpha \beta}+\frac{\theta}{3} h_{\alpha \beta})+4\pi (\tilde
q_\alpha a_\beta+\tilde q_\beta a_\alpha) +4\pi \Pi _{\nu
\beta}(\sigma^\nu_\alpha+\frac{\theta}{3}h^\nu _\alpha)
\label{f}
\end{eqnarray}
\\
\begin{eqnarray}
  E^\lambda_{\mu;\lambda}h^\mu_\alpha+a^\lambda
E_{\alpha \lambda}=-8\pi \tilde
q_\kappa(\sigma^\kappa_\alpha+\frac{\theta}{3}h^\kappa_\alpha)+\frac{4\pi}{3}(2
\tilde \rho
+3\hat P)_{; \nu}h^\nu_\alpha \\ \nonumber
\\ \nonumber
\,\,\,\,\,\,\,\,\,\, \,\,\,\,\,\,\,\,\,\,\,\,\,\,\,\,\,\,\,\,
\,\,\,\,\,\,\,\,\,\, -4\pi(\tilde \rho+\hat P)a_\alpha+4\pi a^\kappa
\Pi_{\alpha \kappa}-4\pi u^{\nu} {\tilde q_{\mu;\nu}}h^\mu_\alpha
\label{g}
\end{eqnarray}

\begin{equation}
E_{\alpha \beta}-4\pi \Pi _{\alpha \beta} =(\frac{3m}{r^3}-4\pi
\tilde \rho)(s_\alpha s_\beta +\frac{1}{3}h_{\alpha \beta})
\label{h}
\end{equation}

In each of equations (\ref{a})--(\ref{h}) there is only one scalar independent component, thus contracting with  $s^\alpha$ we may write the equivalent set:
\begin{equation}
\tilde \rho ^\ast+(\tilde \rho
+\tilde P_r)\theta=\frac{2}{3}(\theta+\frac{\sigma}{2})\Pi-\tilde q^\dagger-2\tilde qa-\frac{2s^1}{r}\tilde q
\label{a'}
\end{equation}

\begin{equation}
\tilde P^{\dagger}_r+(\tilde \rho+\tilde
P_r)a+\frac{2s^1}{r}\Pi=\frac{\sigma}{3}\tilde q-\tilde
q^\ast-\frac{4\theta}{3}\tilde q
\label{b'}
\end{equation}

\begin{equation}
\theta ^\ast+\frac{\theta^2}{3}+\frac{\sigma^2}{6}-a^\dagger
-a^2-\frac{2as^1}{r}=-4\pi(\tilde \rho+3\tilde P_r)+8\pi \Pi
\label{c'}
\end{equation}

\begin{equation}
(\frac{\sigma}{2}+\theta)^\dagger=-\frac{3\sigma s^1}{2r}+12\pi
\tilde q
\label{d'}
\end{equation}

\begin{equation}
E+4\pi \Pi =-a^\dagger -a^2-\frac{\sigma ^\ast}{2}-\frac{\theta
\sigma}{3}+\frac{as^1}{r}+\frac{\sigma^2}{12}
\label{e'}
\end{equation}

\begin{equation}
(4\pi \tilde P_r+\frac{3m}{r^3})(\theta +\frac{\sigma}{2})+(E
-4\pi \Pi +4\pi \tilde \rho)^\ast=-\frac {12\pi s^1}{r}\tilde q
\label{f'}
\end{equation}

\begin{equation}
(E+4\pi\tilde \rho-4\pi
\Pi)^\dagger=\frac{3s^1}{r}(4\pi\Pi-E)+4\pi \tilde
q(\frac{\sigma}{2}+\theta)
\label{g'}
\end{equation}

\begin{equation}
\frac{3m}{r^3}=4\pi \tilde \rho +4\pi (P_\bot -\tilde P_r)+E
\label{h'}                   
\end{equation}
with $f^\dagger=f_{,\alpha}s^\alpha$,
$f^\ast=f_{,\alpha}u^\alpha$ and $a^{\alpha}=as^{\alpha}$, and where the expansion is given by 
\begin{equation}
\theta=u^\mu_{;\mu}=\frac{1}{2(1-\omega^2)^{\frac{1}{2}}}[e^{-\frac{\nu}{2}}(\dot
\lambda+\frac{2\omega \dot \omega}{1-\omega
^2})+e^{-\frac{\lambda}{2}}(\omega \nu ^{\prime}+\frac{2\omega
^{\prime}}{1-\omega ^2}+\frac{4\omega}{r})]
\label{exp}
\end{equation}
 
Then from (\ref{sigma}) and (\ref{exp}) it follows at once that:

\begin{equation}
\frac{\sigma}{2}+\theta=\frac{3\omega s^1}{r}
\label{sigmateta}
\end{equation}
\section{Special cases}
We shall now apply equations (\ref{a'})--(\ref{h'}) to analyze different  particular cases.

\subsection{Geodesic fluids}

If the fluid is geodesic, non--dissipative and locally isotropic, then for bounded configurations, it follows at once from (\ref{b'}) and the vanishing of the pressure at the boundary,
that it should be dust. In this case the vanishing of the Weyl tensor implies the shear free condition as it follows from
(\ref{f'}). On the other hand  the shear free condition implies conformally flat as it follows from (\ref{e'}). Thus in this special case both conditions are equivalent. For
non--geodesic fluids this equivalence is not generally true (see below).

\subsection{Locally isotropic perfect fluids}
Let us now consider locally isotropic and non--dissipative fluids ($\Pi=q=\epsilon=0$) and  find  the relations linking the Weyl tensor,
the shear and the local density inhomogeneity. Although almost all results in this case are known, we think that it is worth while to present them, in order to illustrate the general
method that will be used later to study more complicated situations.

From (\ref{g'}), we obtain after some rearrengements (with $\Pi=\tilde q=0$)
\begin{equation}
[r^3 E]^{\dagger}+r^3 4\pi \rho^\dagger=0.
\label{1g}
\end{equation}
Next, it is convenient to write (\ref{f'}), with the help of (\ref{omega}),  (\ref{a'}), (\ref{h'}), and  (\ref{sigmateta}), as 
\begin{eqnarray}
[r^3 E]^{.}\
+\frac{dr}{dt}[r^3 E]'
=-2\pi
\sigma r^3 (\rho+ P_{r})\sqrt{1-\omega^2}e^{\nu/2}.
\label{npf77n}
\end{eqnarray}

Implying that the vanishing of the Weyl tensor results in the vanishing of spatial 
gradients of energy density and the shear tensor.

Let us now assume
$\rho^\dagger=0$,then  we obtain from
(\ref{1g}) 
\begin{equation}
[r^3 E]^{\dagger}=0.
\label{pf6bis}
\end{equation}
Implying, since the Weyl 
tensor should be regular inside the fluid distribution, 
$E=0$. 
Thus $E=0$ and $\rho^\dagger=0$ are equivalent, and 
either one of them implies $\sigma=0$. These results were already known (see \cite{Lake} and references therein)

Next, if $\sigma=0$ it follows from (\ref{d'}) that
\begin{equation}
\theta^\dagger=0.
\label{pf4}
\end{equation}

Observe that from the above it follows, using equation (\ref{27'}), that if the
fluid is conformally flat and satisfies a barotropic equation of state of
the form $P_{r}=P_{r}(\rho)$, then the fluid is geodesic ($a^{\alpha}=0$).

Also, assuming the shear--free condition alone ($\sigma=0$) it follows from (\ref{npf77n}) that the convective derivative of $E 
r^3$ vanishes, which in turn means that such quantity remains 
constant for any fluid element
along the fluid lines.

\subsection{Locally anisotropic non--dissipative fluids}
We shall now relax the condition of local isotropy of the pressure, 
and shall assume  $\Pi \neq 0$.
Then from (\ref{sigmateta}) and  (\ref{g'}), it follows that 
\begin{equation}
[r^3 (E-4\pi \Pi)]^{\dagger}+r^3 4\pi \rho^\dagger=0.
\label{1ga}
\end{equation}

Implying that the vanishing of $E-4\pi \Pi$ results in the vanishing 
of spatial gradients of energy density.

On the other hand if we assume  the vanishing of $\rho^\dagger$, then assuming that all physical variables are 
regular within the fluid
distribution it follows at once that
\begin{equation}
E-4\pi \Pi=0.
\label{npf8}
\end{equation}
Thus $E-4\pi \Pi=0$ and $\rho^\dagger=0$ are equivalent, but neither one of them implies $\sigma=0$.

Therefore if we assume the spacetime to be conformally flat ($E=0$), then the local anisotropy produces inhomogeneity in the energy density according to the equation
\begin{equation}
(r^3 \Pi)^\dagger= r^3 \rho^\dagger .
\label{npf9bis}
\end{equation}

Next,  it follows from (\ref{f'}), with the help of (\ref{omega}),  (\ref{a'}), (\ref{h'}) and (\ref{sigmateta}) that
\begin{equation}
[r^3(E-4\pi \Pi)]^{.}\
+\frac{dr}{dt}[r^3 (E-4\pi \Pi)]'
+8\pi \Pi\omega e^{(\nu-\lambda)/2} r^2=-2\pi
\sigma r^3 (\rho+P_{r})\sqrt{1-\omega^2}e^{\nu/2}.
\label{npf77a}
\end{equation}
implying thereby that the convective derivative of $r^3(E-4\pi \Pi)$ is controlled not only by $\sigma$, but also by $\Pi$.

If $E=4\pi \Pi$, then the following link between the shear
and  the anisotropy results
\begin{equation}
4\Pi \omega e^{-\frac{\lambda}{2}}=-\sigma r \sqrt{1-\omega^2}(\rho+P_{r}).
\label{npf11}
\end{equation}

 If the fluid is shear-free and $E=4\pi \Pi$ then it is either 
static or locally isotropic. Of course in this last case the fluid is 
also conformally flat.

\subsection{Locally isotropic dissipative fluids in the quasi--static evolution}
We shall now relax the condition of non--dissipation by allowing $q\neq 0$ (for simplicity we put $\epsilon=0$), but assuming that the evolution is slow, which 
means that $\omega^2=\dot
\omega=\dot{ \tilde q} =\ddot \lambda=\ddot \nu=0$ and $q \approx O(\omega)$ (see \cite{HJR}). Then from (\ref{sigmateta}) and (\ref{g'}), we obtain (in the 
quasi--static approximation)
\begin{equation}
E'e^{-\lambda/2}+\frac{3Ee^{-\lambda/2}}{r}=-4\pi
\rho^\dagger ,
\label{df6}
\end{equation}
taking into account that in the quasi--static approximation
\begin{equation}
\rho^{\dagger}=\rho^{\prime}e^{-\frac{\lambda}{2}},
\label{d7}
\end{equation}
we have
\begin{equation}
E^{'}+\frac{3E}{r}=-4\pi \rho^{\prime}.
\label{d8}
\end{equation}

Next,
it follows from (\ref{f'}), with the help of (\ref{a'}) 
\begin{equation}
\dot E
e^{-\nu/2}+E' \omega
e^{-\lambda/2}+\omega e^{-\lambda/2}\frac{3E}{r} -4\pi 
e^{-\frac{\lambda}{2}} {q^{\prime}} +4\pi q 
e^{-\frac{\lambda}{2}}\left(\frac{1}{r}-\nu^{\prime}\right)
=-2\pi
\sigma (\rho+ P_{r}),
\label{d9}
\end{equation}
or, equivalently
\begin{equation}
\frac{e^{\lambda/2}}{4 \pi r^3}(r^3 E)^{\ast} = { 
q^{\prime}} + q \left(\nu^{\prime}-\frac{1}{r}\right) 
-\frac{1}{2}e^{\lambda/2}
\sigma (\rho+ P_{r}).
\label{d10}
\end{equation}

 From (\ref{d8}) it follows at once that conformally flat and 
$\rho^{\prime}=0$ are equivalent conditions, which is also true in 
the perfect (non--dissipating) fluid,
in the quasi--static approximation.

Also, from (\ref{d9}) or (\ref{d10}) it follows that $E=0$ does not 
implies shear--free. Indeed,assuming $E=0$ in (\ref{d10}) we have
\begin{equation}
  q^{\prime} + q \left(\nu^{\prime}-\frac{1}{r}\right) 
=\frac{1}{2}e^{\lambda/2}
\sigma (\rho+P_{r}),
\label{d11}
\end{equation}
yielding
\begin{equation}
  q=r e^{-\nu}\left[ \int \frac{ e^{\lambda/2+\nu}}{2 r}
\sigma (\rho+ P_{r})dr +\beta(t)\right].
\label{d12}
\end{equation}
If,  we further impose the shear--free condition then, one obtains 
from (\ref{d12})
\begin{equation}
q=r \beta (t)e^{-\nu}
\label{d13}
\end{equation}
leading to a condition on the
temperature, which may be obtained using the Landau-Eckart equation
\begin{equation}
 q_{\mu}=\kappa h^{\nu}_{\mu}(T_{,\nu}-T a_{\nu}),
\label{d14}
\end{equation}
or
\begin{equation}
 q=-\kappa e^{-\frac{\lambda}{2}}(T^{'}+\frac{T \nu^{\prime}}{2}).
\label{d15}
\end{equation}
Using (\ref{d15}) in (\ref{d13}) we obtain
\begin{equation}
T=e^{-\nu/2}\left[C(t)-\frac{\beta(t)}{\kappa}\int^{r}_{0}re^{(\lambda-\nu)/2} dr\right].
\label{d16}
\end{equation}
The two functions $\beta(t)$ and $C(t)$ are simply related to the 
total luminosity of the sphere and the central 
temperature, through (\ref{d13}) and
(\ref{d16}), respectively.
A simple model satisfying $E=\sigma=0$ wil be next presented.
\subsection{A conformally flat, shear--free sphere, dissipating in the quasi--static regime (with $\epsilon=0$)}
From (\ref{3'}) and $E=0$ we have 
\begin{equation}
8\pi P_{\perp}=\frac{e^{-\lambda}}{r}\left(\nu^{\prime}-\lambda^{\prime}-\frac{1}{r}\right)+
\frac{1}{r^2},
\label{7}
\end{equation}
Then substracting (\ref{2}) from (\ref{7}), and considering $\Pi=0$, we obtain
\begin{equation}
e^{-\lambda}=r^2c_1+1,
\label{9'}
\end{equation}
where $c_1(t)$ is an arbitrary function of time.
Substituting   (\ref{9'}) into (\ref{1}) yields
\begin{equation}
8\pi \rho=-3c_1.
\label{10}
\end{equation}

Considering $E=0$ with (\ref{9'}) we obtain
\begin{equation}
\nu^{\prime\prime}+\frac{\nu^{\prime 2}}{2}-\frac{\nu^{\prime}}{r(r^2c_1+1)}=0,
\label{11}
\end{equation}
which has the solution
\begin{equation}
e^{\nu/2}=(r^2c_1+1)^{1/2}c_2+c_3,
\label{12'}
\end{equation}
where $c_2(t)$ and $c_3(t)$ are arbitrary functions of $t$.

Substituting (\ref{9'}) and (\ref{12'}) into (\ref{2}) we obtain
\begin{equation}
8\pi P_{r}=c_1\left[1+2c_2e^{-(\lambda+\nu)/2}\right].
\label{13}
\end{equation}

From (\ref{sigma}), condition  $\sigma=0$ can be rewriten as 
\begin{equation}
\left(\frac{\omega e^{\nu/2}}{r}\right)^{\prime}=-\frac{\dot{\lambda}e^{\lambda/2}}{2r},
\label{14}
\end{equation}
which after integration becomes
\begin{equation}
\omega=\left(c_4-\frac{\dot{c}_1}{2c_1}e^{\lambda/2}\right)re^{-\nu/2},
\label{15}
\end{equation}
where $c_4(t)$ is an arbitrary function of $t$.

From (\ref{9'}), (\ref{10}), (\ref{12'}), (\ref{13}) we have
\begin{equation}
8\pi(\rho+P_{r})\omega=-\left(2c_1c_4-\dot{c}_1e^{\lambda/2}\right)rc_3e^{-\nu}.
\label{16'}
\end{equation}
Now substituting (\ref{9'}), (\ref{12'}), (\ref{16'}) into (\ref{4}) we obtain
\begin{equation}
8\pi q=\left(2c_1c_3c_4+\dot{c}_1c_2\right)re^{-\nu}.
\label{17}
\end{equation}
Next, using  (\ref{enusigma}) and (\ref{elambdasigma})  in (\ref{9'}) and (\ref{12'}) we obtain
\begin{equation}
c_{1}=-\frac{2M}{r_{\Sigma}^3},
\label{c1}
\end{equation}
and 
\begin{equation}
c_{3}=\sqrt{1-\frac{2M}{r_{\Sigma}}}(1-c_{2}).
\label{c2p}
\end{equation}
Also, from the junction condition (\ref{PQ})
and from (\ref{15}) evaluated at the boundary surface, it follows that
\begin{equation}
c_{1}(1+2c_{2})=(2c_{1}c_{3}c_{4}+\dot c_{1}c_{2})\frac{r_{\Sigma}}{(1-\frac{2M}{r_{\Sigma}})},
\label{j1}
\end{equation}
and
\begin{equation}
\dot c_{1}=2c_{1}\sqrt{1-\frac{2M}{r_{\Sigma}}}\left(c_{4}-\frac{\omega_{\Sigma}}{r_{\Sigma}}\sqrt{1-\frac{2M}{r_{\Sigma}}}\right).
\label{j2}
\end{equation}
Solving algebraically the system (\ref{c2p})--(\ref{j2}) for $c_{2}, c_{3}$ and $c_{4}$, we can express these functions in terms of $M, r_{\Sigma}, \omega_{\Sigma}$ and $\dot
M$.

We shall further specify our model by assuming
\begin{equation}
c_{2}=1,
\label{c2}
\end{equation}
\begin{equation}
c_{3}=0,
\label{c3}
\end{equation}
implying
\begin{equation}
e^{-\lambda}=e^{\nu}=1-\frac{2Mr^2}{r_{\Sigma}^3}.
\label{metrica}
\end{equation}
Using (\ref{c1}), (\ref{c2}), (\ref{c3}) and (\ref{metrica}) in (\ref{10}) and
(\ref{13}) we obtain
\begin{equation}
8\pi \rho=-8\pi P{_r}=\frac{6M}{r_{\Sigma}^3}.
\label{pi}
\end{equation}
Then using  (\ref{c2}) and (\ref{c3}) in (\ref{17})
we obtain
\begin{equation}
8\pi q=r\dot c_{1}e^{-\nu},
\label{q}
\end{equation}
which is our equation (\ref{d13}) with $\frac{\dot c_{1}}{8\pi}=\beta (t)$. By virtue of (\ref{c1}) and (\ref{j1}), the expression for $q$ becomes
\begin{equation}
8\pi q=-\frac{6Mr}{r_{\Sigma}^4} \frac{(1-\frac{2M}{r_{\Sigma}})}{(1-\frac{2Mr^2}{r_{\Sigma}^3})}.
\label{qn}
\end{equation}
Next, using (\ref{c1}), (\ref{j1}), (\ref{metrica}) in (\ref{d16}), we obtain
\begin{equation}
T=\frac{1}{\sqrt{1-\frac{2Mr^2}{r_{\Sigma}^3}}}\left[T_{c}-\frac{3}{16\pi \kappa r_{\Sigma}}(1-\frac{2M}{r_{\Sigma}})\log{(1-\frac{2Mr^2}{r_{\Sigma}^3}})\right],
\label{d17}
\end{equation}
with $T_{c}$ denoting the temperature at $r=0$, and from (\ref{15})
\begin{equation}
\omega=\frac{r}{r_{\Sigma}}\sqrt{\frac{1-\frac{2M}{r_{\Sigma}}}{1-\frac{2Mr^2}{r_{\Sigma}^3}}}\left[\frac{3}{2}\left(1-\sqrt{\frac{1-\frac{2M}{r_{\Sigma}}}{1-\frac{2Mr^2}{r_{\Sigma}^3}}}
\right)+\omega_{\Sigma}\right].
\label{omegan}
\end{equation}

Thus our model represents a conformally flat  sphere of fluid evolving slowly and shear free, with homogeneous energy density and pressure, satisfying the ``inflationary'' equation of
state $\rho+ P_{r}=0$, with an inward ($q<0$) heat flux.

\subsection{General case}
First of all, it should be noticed that in the presence of, both,  anisotropic pressure and dissipation, for $B_{1}$  warped product spacetimes (which include the spherically
symmetric case) there exist some restrictions on physical variables, provided by the syzygy \cite{lake}. However in our  case, as it follows from the definitions in subsection
(2.2) we have $\mid
\tilde q^{\nu}
\tilde q_{\nu}
\mid =(s_{\nu} \tilde q^{\nu})^{2}$ (notice that our signature is $-2$), and therefore the syzygy (see equation (37) in \cite{lake}) reduces to the identity $0=0$.

Then, we obtain  from (\ref{sigmateta}) and (\ref{g'}), after some rearrengements, 
\begin{equation}
[r^3 (E-4\pi \Pi-4\pi \tilde q \omega)]^{\dagger}+r^3 4\pi \tilde \rho^\dagger=-4 \pi(\tilde q \omega)^{\dagger}r^3 ,
\label{1gg}
\end{equation}
or, equivalently
\begin{equation}
[r^3 (E-4\pi \Pi)]^{\dagger}+4\pi r^3 \tilde \rho^\dagger=4 \pi r^3 \tilde q (\frac{\sigma}{2}+\theta),
\label{1ggbis}
\end{equation}
if $\tilde q=0$ we recover (\ref{1ga}).

Also, from (\ref{f'}), with the help of (\ref{a'})  we get
\begin{eqnarray}
[r^3(E-4\pi \Pi)]^{.}\
+\frac{dr}{dt}[r^3 (E-4\pi \Pi)]'
+8\pi \Pi\omega e^{(\nu-\lambda)/2} r^2 \nonumber \\
+4\pi r^3 \sqrt{1-\omega^2}e^{\nu/2}[\tilde q(\frac{s^1}{r}-2a)-\tilde q^{\dagger}]=-2\pi
\sigma r^3 (\tilde \rho+\tilde P_{r})\sqrt{1-\omega^2}e^{\nu/2},
\label{npf77nn}
\end{eqnarray}
which yields (\ref{npf77a}) in the non--dissipative case.

From (\ref{1ggbis}) it is clear that the appearance of inhomogeneities in the energy--density are controlled by the Weyl tensor, the anisotropy of the pressure and the dissipation. If
we express the heat flux through the temperature, the relaxation time and the heat conduction coefficient, by means of some transport equation (e.g.
\cite{Israel}) then the aforementioned parameters may be related with the creation of such density inhomogeneities.

Indeed, assuming for example the transport equation \cite{maartens}
\begin{equation}
\tau h^\mu_\nu  (q^\nu)^\ast + q^\mu = \kappa h^{\mu\nu} (T_{,\nu} - T
a_\nu) - \frac{1}{2} \kappa T^2 \left(\frac{\tau u^\alpha}{\kappa
T^2}\right)_{;\alpha} q^\mu 
\label{tr}
\end{equation}
where $\tau, \kappa$, and $T$  denote the relaxation time,
the thermal conductivity and  the temperature,
respectively. We obtain, putting for simplicity $\epsilon=0$, and using (\ref{b'})

\begin{equation}
q=\frac{\tau\left[P_r^{\dagger} + (\rho + P_r) a + \frac{2}{3 \omega}\left(\frac{\sigma}{2}+\Theta\right) \Pi\right] - \kappa \left(T^{\dagger} + T a \right)}{1+ \frac{\tau}{2}
\left[\frac{1}{3} \left(2\sigma - 5\Theta\right)  +
\frac{\tau^{\ast}}{\tau} - \frac{\kappa^{\ast}}{\kappa} - \frac{2 T^\ast}{T}\right]}
\label{qtau}
\end{equation}
Then replacing (\ref{qtau}) into (\ref{1ggbis}) (with $\epsilon =0$) we obtain an expression which brings out the role of thermodynamic variables in the appearance of density
inhomogeneities, namely
\begin{eqnarray}
&&[r^3 (E-4\pi \Pi)]^{\dagger} + 4\pi r^3 \tilde \rho^\dagger \nonumber \\
&=&4 \pi r^3 (\frac{\sigma}{2}+\theta) \frac{\tau\left[P_r^{\dagger} + (\rho + P_r) a + \frac{2}{3
\omega}\left(\frac{\sigma}{2}+\Theta\right) \Pi\right] - \kappa \left(T^{\dagger} + T a \right)}{1+ \frac{\tau}{2}
\left[\frac{1}{3} \left(2\sigma - 5\Theta\right)  +
\frac{\tau^{\ast}}{\tau} - \frac{\kappa^{\ast}}{\kappa} - \frac{2 T^\ast}{T}\right]},\nonumber \\
\label{1ggbisbis}
\end{eqnarray}

A similar conclusion applies to the shear of the fluid, as it follows from (\ref{npf77nn}).

With the sole purpose of bringing  out the role of dissipation in the formation of density inhomogeneities, let us present a simple and highly idealized model.

\subsection{A dissipative model with $E-4\pi \Pi =0$}

Since we want to exhibit the role of dissipation in the formation of ihomogeneities we shall assume $E-4\pi \Pi =0$, then from (\ref{rieman}) and (\ref{66t}) it follows that

\begin{equation}
E-4\pi \Pi=0\Leftrightarrow m=\frac{4\pi}{3}\tilde \rho
r^3\Leftrightarrow e^{-\lambda}=1-\frac{8\pi}{3}\tilde \rho r^2
\label{3}
\end{equation}
From (\ref{f'}), (\ref{1ggbis}) and (\ref{sigmateta}) we obtain 
\begin{equation}
\tilde \rho^\dagger=\tilde q\frac{3\omega s^1}{r},
\label{9}
\end{equation}
and
\begin{equation}
3s^1r^2(\tilde q+\tilde P_r \omega)+(\tilde \rho r^3)^\ast=0,
\label{8}
\end{equation}
then combining the Einstein equations (\ref{fieq00}) and (\ref{fieq11}) with (\ref{3}), it follows that
\begin{equation}
\nu ^{\prime}=\frac{8\pi(\tilde \rho+\tilde \rho ^\prime r+3\tilde
P_r)r}{3(1-\frac{8\pi}{3}\tilde \rho r^2)}.
\label{11}
\end{equation}

We shall further assume the equation of state
\begin{equation}
\tilde P_r=\frac{1}{3}\tilde \rho
\label{23}
\end{equation}
then, replacing (\ref{23}) into (\ref{11}) we obtain
\begin{equation}
e^\nu=\frac{(1-\frac{8\pi}{3}\tilde \rho _\Sigma
r^2_\Sigma)^2}{1-\frac{8\pi}{3}\tilde \rho r^2}.
\label{24}
\end{equation}
Next, using (\ref{8}) and (\ref{9}) it follows that:
\begin{equation}
\tilde \rho ^\dagger =\frac{3\omega \tilde q}{r}\sqrt{\frac{1-\frac{8\pi}{3}\tilde \rho r^2}{1-\omega^2}} 
\label{26}
\end{equation}
and
\begin{equation}
\omega=\frac{-\dot{\tilde \rho}r \pm \sqrt{(\dot{\tilde
\rho}r)^2-r\tilde \rho ^{\prime}(r\tilde \rho ^{\prime}+4\tilde
\rho)(1-\frac{8\pi}{3}\tilde \rho_\Sigma r^2 _\Sigma)^2}}{(r\tilde
\rho ^{\prime}+4\tilde \rho)(1-\frac{8\pi}{3}\tilde \rho_\Sigma
r^2 _\Sigma)}
\label{27}
\end{equation}
In the particular  case $\tilde \rho=\tilde \rho (t)$ (but $\tilde \rho ^\dagger \neq 0$), these last equations reduce to
\begin{equation}
\omega=\frac{-\dot{\tilde \rho}r }{2\tilde
\rho(1-\frac{8\pi}{3}\tilde \rho_\Sigma r^2 _\Sigma)}
\label{28}
\end{equation}
and
\begin{equation}
\tilde q =\frac{1}{3}\dot{\tilde \rho}re^{\frac{\lambda-\nu}{2}}
\label{21}
\end{equation}

From junction conditions it is very simple to express $\tilde \rho$ and $\dot{\tilde \rho}$, and thereof all physical and metric variables in terms of the total mass $M$, the velocity
of the surface
$\omega_{\Sigma}$, the radius
$r_{\Sigma}$ and the luminosity of the sphere.
Equation (\ref{26}) shows how dissipation produces density inhomogeneity. It is worth noticing that both kinds of dissipative processes ($\epsilon$ and $q$) may produce such
inhomogeneity.
\section{Conclusions}
The study of general relativistic gravitational collapse, which has attracted
the attention of researchers since  the seminal paper by Oppenheimer
and Snyder \cite{Opp}, is mainly motivated by the fact that the gravitational collapse of massive stars represents one of
the few observable phenomena, where general relativity is expected to
play a relevant role. Ever since that work, much was written by
researchers trying to provide models of evolving gravitating spheres.
However this endeavour proved to be difficult and uncertain. Different
kinds of advantages and obstacles appear, depending on the approach adopted
for the modeling.

Here, we have established the set of equations governing the structure and evolution of self--gravitating spherically symmetric dissipative anisotropic fluids. For reasons
explained in the Introduction, emphasis has been put on the role played by the Weyl tensor, the anisotropy of the pressure, dissipation, density inhomogeneity and the shear tensor.

 The particular  simple relation between the Weyl tensor and
density inhomogeneity, for perfect fluids  (also valid for locally isotropic, dissipative fluids in the quasistatic regime),  is at the origin of the Penrose's  proposal to provide a
gravitational arrow of time. However the fact that such relationship is no longer valid in the presence of local anisotropy of the pressure and/or dissipative processes, explains its
failure in scenarios where the abovementioned  factors are present.

From (\ref{1ggbis}) it is apparent that  the production of density inhomogeneities is related to a quantity involving all those factors ($[r^3 (E-4\pi \Pi)]^{\dagger}-4 \pi r^3 \tilde
q (\frac{\sigma}{2}+\theta)$). Alternatively, the appearance of such inhomogeneities may be described by means of thermodynamical variables (if a transport equation is assumed), as
indicated in (\ref{1ggbisbis}).

This situation is illustrated in the example provided in the last section, where density inhomogeneity is produced by dissipative processes alone. Thus if
following Penrose we adopt the point of view that self--gravitating systems evolve in the sense of increasing of density inhomogeneity, then the absolute value of the quantity  above
(or some function of it) should increase, providing an alternative definition for an arrow of time.

\section{Acknowledgements}
J.M and J.O  acknowledge financial assistance under grant
BFM2003-02121 (M.C.T. Spain).

\end{document}